\documentclass[final]{IEEEtran}
\pdfoutput=1
\IEEEoverridecommandlockouts
\usepackage{cite}
\usepackage{amsmath,amssymb,amsfonts}
\usepackage{algorithmic}
\usepackage{graphicx}
\usepackage[update,prepend]{epstopdf}
\usepackage{textcomp}
\usepackage{xcolor}
\usepackage{multirow}
\usepackage{placeins}
\usepackage{bbm}
\usepackage{color} 
\usepackage{soul} 
\usepackage{algorithm2e}
\usepackage{booktabs}
\def\BibTeX{{\rm B\kern-.05em{\sc i\kern-.025em b}\kern-.08em
    T\kern-.1667em\lower.7ex\hbox{E}\kern-.125emX}}

\begin{document}
\setlength{\parskip}{0pt}
\title{Linear Jamming Bandits: Learning to Jam 5G-based Coded Communications Systems  
\thanks{This work was supported by the Department of Defense Cyber Scholarship Program.}}

\author{\IEEEauthorblockN{
     Zachary Schutz\IEEEauthorrefmark{1},
     Daniel J. Jakubisin\IEEEauthorrefmark{1},
     Charles E. Thornton\IEEEauthorrefmark{1},
     R. Michael Buehrer\IEEEauthorrefmark{2}   
} \\
\IEEEauthorblockA{
     \IEEEauthorrefmark{1}Virginia Tech National Security Institute, Blacksburg, VA (Email: zachschutz27@vt.edu; djj@vt.edu) \\ 
     \IEEEauthorrefmark{2}Wireless@VT, Bradley Department of ECE, Virginia Tech, Blacksburg, VA
    } }

\maketitle

\begin{abstract}
We study jamming of an OFDM-modulated signal which employs forward error correction coding. We extend this to leverage reinforcement learning with a contextual bandit to jam a 5G-based system implementing some aspects of the 5G protocol. This model introduces unreliable reward feedback in the form of ACK/NACK observations to the jammer to understand the effect of how imperfect observations of errors can affect the jammer's ability to learn. We gain insights into the convergence time of the jammer and its ability to jam a victim 5G waveform, as well as insights into the vulnerabilities of wireless communications for reinforcement learning-based jamming.
\end{abstract}
\begin{IEEEkeywords}
Reinforcement Learning, Linear Bandits, Jamming, Thompson Sampling, 5G, Forward Error Correction
\end{IEEEkeywords}

\section{Introduction} \label{Intro}

Adaptation of a jammer's waveform to the victim's communication signal can significantly increase the effectiveness and efficiency of the jammer. 
However, prior information about the victim system must be known such as the waveform used and the frequency band it occupies. 
Reinforcement learning (RL) can be leveraged in this situation because it does not need prior information in order to learn to effectively jam a victim signal.
The studies of these issues also often assume the jammer has perfect observations of the errors caused to the victim signal, but in reality, this is not the case. There is a need to study the behavior of the jammer under RL when reward feedback is unreliable to observe if the jammer is able to converge to a meaningful solution with no prior information.  

Previous works explored using RL in a bandit setting to jam communications systems \cite{Amuru2016,Thornton2022,Schutz2024}. 
One assumption the previous works make is that the victim signal does not apply forward error correction (FEC) coding.
FEC adds redundancy to the data stream to detect and correct errors in the bit stream due to noise, fading, and multipath propagation. The redundancy in the data lowers throughput, but ensures the data is sent correctly over the communications channel by achieving both time and frequency diversity. This introduces the need to examine performance of the bandit against a more realistic victim signal.

The jamming rate and power scaling parameter, \(\rho\), was used to increase the instantaneous power of the jammer to exploit the structure of the victim signal while keeping the average power the same across a moment in time \cite{Thornton2022,Schutz2024}. This was especially important at higher SNRs where the victim signal had a higher chance of being decoded correctly due to the difference in average signal-to-jammer ratio (SJR) power.

The reward metric in the prior work was the symbol error rate (SER).  However, in the case of communications employing linear block codes, such as 5G, the proper reward metric becomes the block error rate (BLER) of the victim signal.  This changes the impact of the jammer's pulse parameter \(\rho\), which we study in this paper.

A 5G-based system will also employ an uplink and downlink channel with different scheduling techniques that make it difficult for a jammer to know what frequency and time the victim Tx/Rx pair is transmitting on. This in turn makes it difficult for the jammer to have perfect observations of the errors to see how effective its jamming strategies were. By exploring how RL will learn to jam a coded-based system, we hope to gain insight on how the agent will learn
and how to adjust its actions in order to learn fast and efficiently. 
Imperfect feedback observations are thus integrated to demonstrate how the bandit adapts to receiving unreliable information.

For the following sections, we introduce the overall system model to be used for the victim and jamming signals. We then examine optimal jamming against OFDM-modulated signals independent of any learning scheme. The 5G-based system is then introduced, as well as a summary on how to practically jam the victim with unreliable reward information. We then explain how to apply RL to a jammer seeking to disrupt a 5G-based communications system, and lastly, the analysis of jamming the 5G-based system is examined as well as the impact of unreliable reward information on the jammer's ability to learn an effective solution.
\section{System Model}\label{System Model}
We model the legitimate signal between transmitter and receiver (i.e., the ``victim signal'') as an OFDM-modulated symbol. For the victim, the OFDM-modulated signal employs FEC. In the time-domain (TD), the victim signal is represented as
\begin{equation} \label{eq:OFDMsignal}
    v(t) = \sum_{k = 1}^{N_{sc}} V(k) exp(j2\pi kt/N_{sc}), \forall 0\leq t\leq N_{sc}-1
\end{equation}
where \(V(k)\) is the frequency-domain (FD) symbol, that is typically a digital phase-amplitude-modulated signal, such as M-QAM or M-PSK, and \(N_{sc}\) is the total number of subcarriers used in the OFDM transmission. The IFFT operation converts the frequency-modulated symbols to TD signals before they are transmitted. 

The jamming signal employs a multi-carrier waveform. We refer to this type of jamming as frequency-domain (FD) jamming. We do not consider single-carrier or TD jamming schemes, since it was previously shown that TD jamming strategies are suboptimal for jamming OFDM-modulated signals \cite{Schutz2024}. The jamming signal modeled in the TD is
\begin{equation}\label{eq:banditOFDM}
    j(t) = \sum_{k = 1}^{N_{sc}} J(k) exp(j2\pi kt/N_{sc}), \forall 0\leq t\leq N_{sc}-1
\end{equation}
where \(J(k)\) is the FD symbol that is typically a digital phase-amplitude-modulated signal, such as M-QAM or M-PSK, and \(N_{sc}\) is the total number of subcarriers used in the OFDM transmission. The IFFT operation converts the frequency-modulated symbols to TD signals before they are transmitted. 

Here it is assumed that the victim signal, \(v(t)\), is attacked by a jamming signal, \(j(t)\), and is also effected by a zero-mean Gaussian noise term, \(n(t)\), with variance \(\sigma^2\). It is assumed that the noise has constant power over the observation interval. The received signal at the victim's receiver can be expressed as
\begin{equation} \label{eq:RxSignal}
    y(t) = \sqrt{P_v}v(t) + \sqrt{P_j}j(t)exp(j\phi) + n(t)
\end{equation}
where \(P_v\) and \(P_j\) are the power of the victim signal and power of the jamming signal, respectively, and \(\phi \in (0, 2\pi] \) is a uniform random variable representing the jamming signal's phase-offset from the victim signal. 

At the receiver, an FFT operation is performed to return the underlying frequency modulated victim symbols, \(V(k)\),
\begin{align}\label{eq:FDtotalJam}
    &Y(k)\sum_{t = 0}^{N_{sc}-1} y(t)exp(-j2\pi kt/N_{sc}),\forall 0\leq k\leq N_{sc}-1\notag \\
    &= \sqrt{P_v}V(k) + \sqrt{P_j}J(k)exp(j\phi) + N(k),
\end{align}
where \(N(k)\) is the FD version of \(n(t)\) after the FFT.
It is assumed the jamming and victim signals operate in the same frequency band. The victim is not able to 
employ frequency hopping.
The main concern is to observe how the jammer learns to jam a victim employing an OFDM-modulated signal with FEC. 
When the jammer is coherent with the victim signal, the phase term reduces to 1 (i.e., \(\phi = 0\)).
It is assumed for a FD jamming signal that all of its power is focused in the data-carrying subcarriers of the victim signal, and no power is added in the guard band or cyclic prefix. 

The average power of the jamming signal as measured by the jammer-to-noise ratio (JNR) is seen as constant at the victim's receiver, but the instantaneous power may be increased by implementing a pulsed jammer. In this case, the jamming signal is modified by \(\rho\) where \(\rho\) is the probability the jammer is on and \((1-\rho)\) is the probability the jammer is off. The term \(\rho\) modifies the instantaneous power by
\begin{equation}
    P_{j_{inst}} = \frac{10^{JNR/10}}{\rho}
\end{equation}
where we assume \(\sigma^2 = 1\) for the purposes of setting \(P_j\) and \(P_v\).

\section{Analysis of Jamming FEC Coded OFDM} \label{sec:OFDMcodeJamming}
We do not consider RL for this section to strictly observe how to best jam an OFDM-modulated signal employing FEC. 
The code uses soft-decision decoding in order to decode the bits. The jammer generates an AWGN signal modified by \(\rho\) to examine how different values for \(\rho\) at a constant SNR and JNR affect the victim signal. We consider phase coherence between the jammer and victim. Both signals are also sent through an AWGN channel, as well as being affected by the jammer producing an AWGN signal.

\begin{figure}[ht]
	\centerline{\includegraphics[width=0.65\columnwidth]{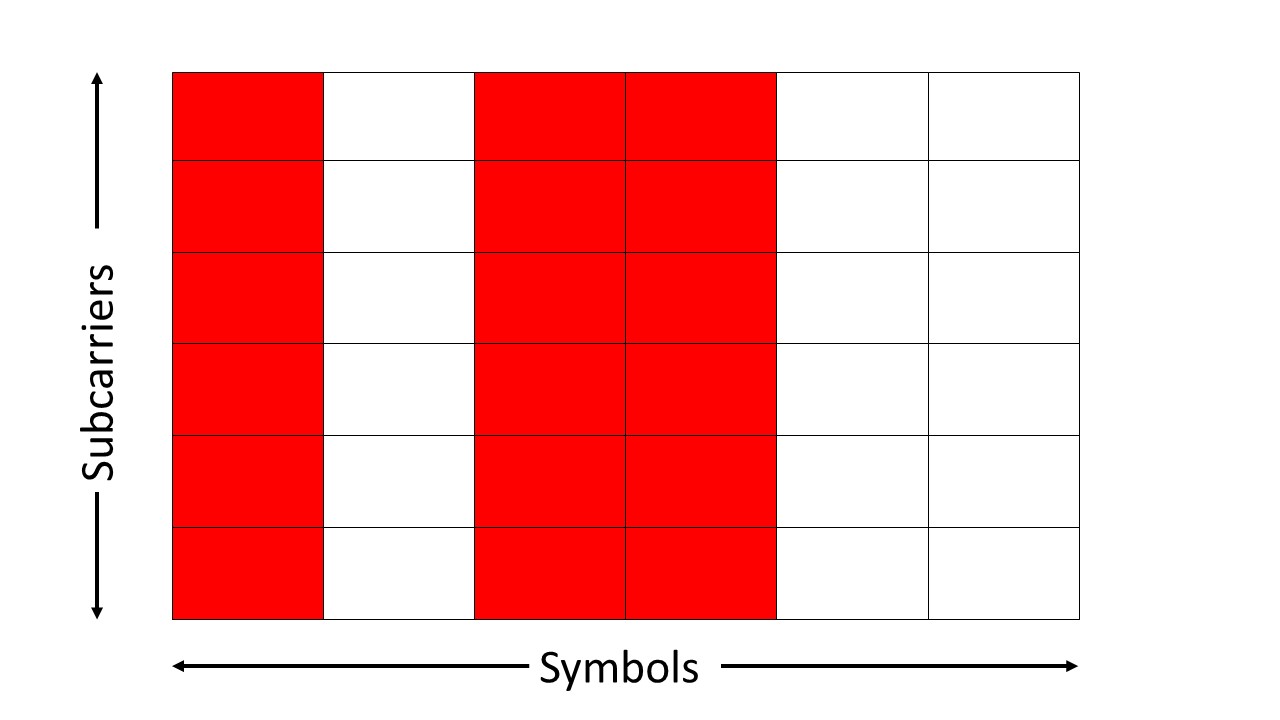}}
	\caption{Example of symbol jamming where \(\rho\) distributes power over time.}
	\label{fig:ExampleSymJam}
\end{figure}

For this section, the victim signal employs an OFDM-modulated signal with BPSK modulation on the subcarriers with a 3/4 rate LDPC code with a block size of 27. The OFDM signal consists of 324 data subcarriers, 83 total guard bands, 41 guard bands on each end of the signal and a null on the center subcarrier, and a cyclic prefix length of 27. There are 2 codewords per OFDM symbol. We examine the effect of using 
FD AWGN to jam the signal where it is able to focus its energy in the data-carrying subcarriers.
We also implement 3 types of jamming: symbol, subcarrier, and randomly jamming the resource elements within the resource grid, which we will call random jamming. Symbol jamming applies the jammer on an OFDM symbol basis. When \(\rho\) is applied, it is on a symbol-by-symbol basis and adjusts power by jamming the subset of selected symbols at a higher instantaneous power, as seen in Fig. \ref{fig:ExampleSymJam}. Subcarrier jamming applies the jammer on a subcarrier basis only. The parameter \(\rho\) is applied on a subcarrier-by-subcarrier basis to the selected subcarriers where the power is distributed, and the non-selected subcarriers are not affected at all.
Random jamming pulses the jammer within the subcarrier-symbol grid.
These jamming types are examined such that they are the only jamming type used for a simulation to compare and contrast how jamming different parts of an OFDM signal may affect the block errors achieved by the jammer.

\begin{figure}[ht]
	\centerline{\includegraphics[width=1.0\columnwidth]{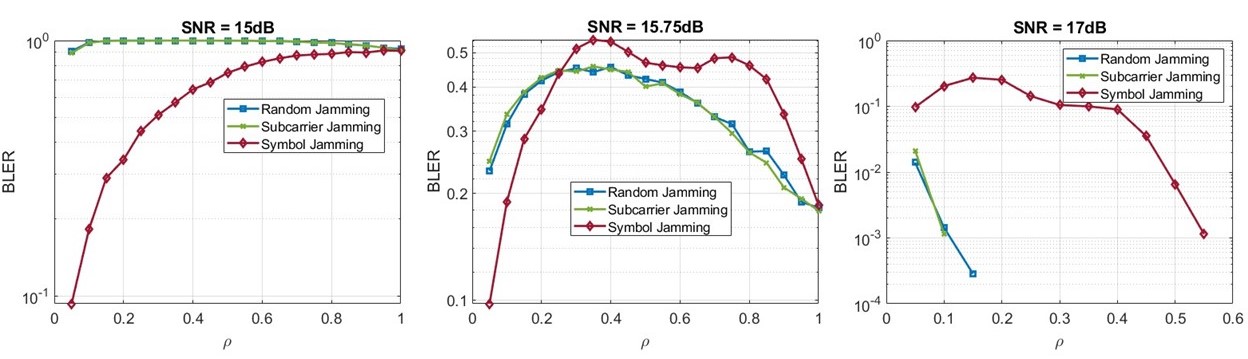}}
	\caption{Average BLER across \(\rho\) using different jamming methods at JNR = 10 dB. Increasing SNR is displayed from left to right: 15 dB, 15.75 dB, and 17 dB.}
	\label{fig:BLERFD}
\end{figure}

The BLER simulations over \(\rho\) shown in Fig. \ref{fig:BLERFD} are of SNR values of 15 dB to 17 dB. 
We observed that all jamming types achieve BLERs of nearly 100\% below an SNR of 15 dB, and above an SNR of 17 dB, the jammer will achieve almost no block errors, so we do not show these cases. Intermediate values such as SNR = 15.75 dB were also included to show the sensitivity of BLER to SNR.

Initially at an SNR of 15 dB, the jamming methods of random and subcarrier jamming perform the best since at any value for \(\rho\) a high BLER is achieved. Symbol jamming does not perform well at lower values for \(\rho\) since it does not hit enough codewords to have a large effect. 

As SNR increases to 15.75 dB, the BLER returns for random and subcarrier jamming diminish quickly compared to 15 dB, while symbol jamming is able to outperform the other jamming methods. 
The sharp inflection for this case can be explained by how the codewords are placed in the subcarrier-symbol grid and the jamming rate used. Codewords are mapped across subcarriers for a specific OFDM symbol. Thus, symbol jamming concentrates the jammer's power into a subset of the transmitted codewords. 
Symbol jamming has subpar performance at extreme \(\rho\) values because it either does not hit enough codewords to have an effect or it does not have enough instantaneous power to effect the codewords. The middle values for \(\rho\) therefore have the best combination of jamming rate and instantaneous power to effect the codewords.  

Increasing the SNR to 17 dB, random and subcarrier jamming have enough instantaneous power to have only minimal effects on the codewords. Random and subcarrier jamming 
will only hit parts of codewords while symbol jamming is guaranteed to hit two codewords per OFDM symbol. It is clear in this caee that with symbol jamming that low \(\rho\) values perform the best.


\begin{figure}[ht]
	\centerline{\includegraphics[width=1.0\columnwidth]{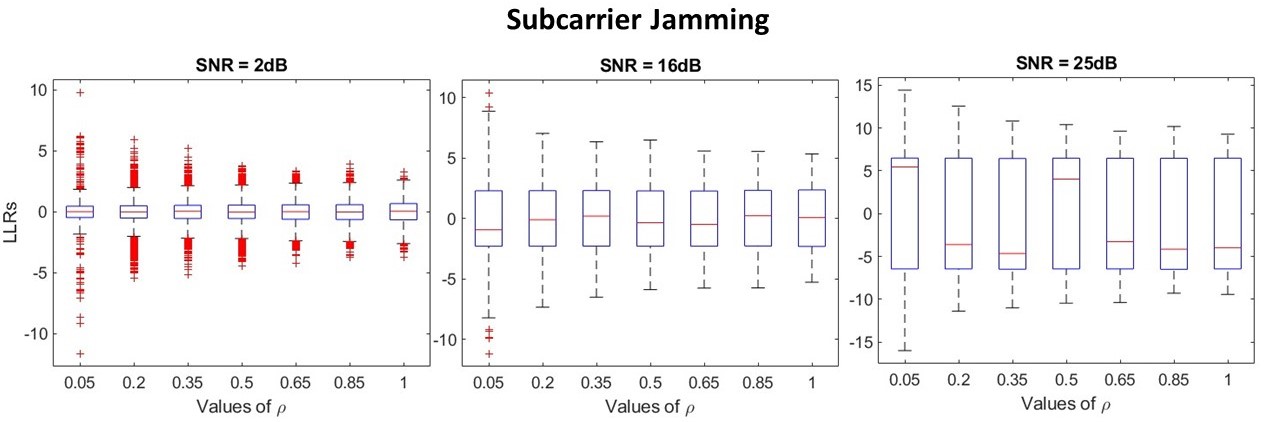}}
	\caption{Box plots of LLRs across \(\rho\) using subcarrier jamming and FD AWGN with JNR = 10 dB.}
	\label{fig:BoxSubOFDMFD}
\end{figure}

\begin{figure}[ht]
	\centerline{\includegraphics[width=1.0\columnwidth]{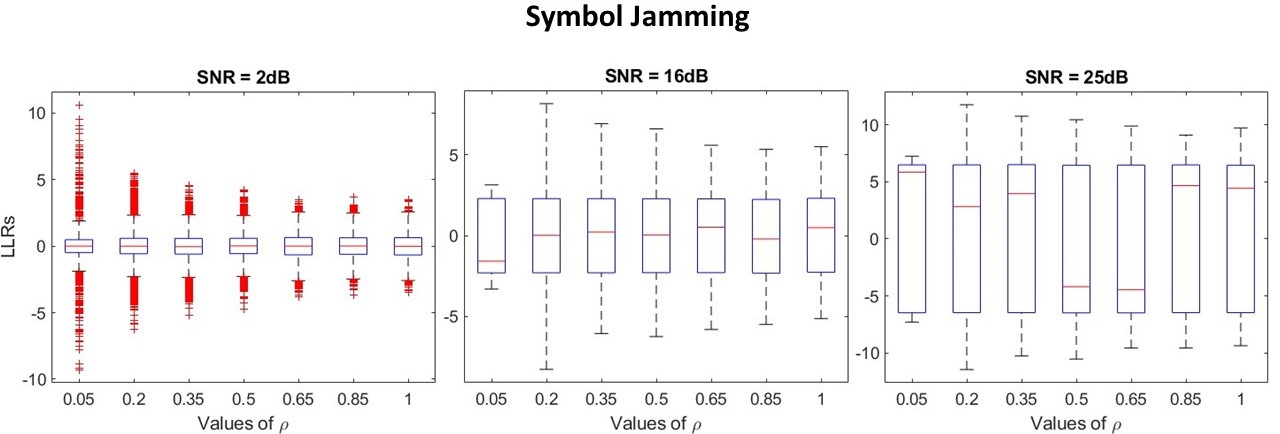}}
	\caption{Box plots of LLRs across \(\rho\) using symbol jamming and FD AWGN with JNR = 10 dB.}
	\label{fig:BoxSymbOFDMFD}
\end{figure}

Figs. \ref{fig:BoxSubOFDMFD} and \ref{fig:BoxSymbOFDMFD} show the LLR distributions over \(\rho\) as box plots for subcarrier jamming and symbol jamming.
We omit random jamming, which had the same performance as subcarrier jamming.
For random and subcarrier jamming, as \(\rho\) increases, the spread of the distribution and number of outliers decrease. For symbol jamming, the distribution is initially small, but as \(\rho\) increases, the distribution increases and then shrinks again. This helps support our statements made above of the BLER curves by showing a similar trend in LLR distribution. 

A typical LTE or 5G BLER is 2\% while BLERs greater than or equal to 10\% are unacceptable for successful transmissions. According to these simulations, with a jammer operating at a JNR of 10 dB present, an SNR above 17 dB would be needed to achieve successful transmissions.
\subsection{Insights on Jamming Coded Signals}
The BLER has unique results depending on the jamming method and \(\rho\) value used.
For low SNR, high values of \(\rho\) appear to be the most useful. For high SNR, low \(\rho\) is better to focus energy into certain codewords or parts of codewords to guarantee a block error.
The above analysis gives insight in how jamming a coded system might work, but what is unique to these cases is how the codeword is layed out and how many codewords are in each symbol. The same results might not be achieved if one codeword was dedicated to one subcarrier-symbol block. These results are still useful because targeting whole codewords instead of parts of codewords seem to be the most effective in causing transmission errors.

\section{Jamming a 5G-Based System}
We now introduce an RL bandit to jam a 5G-based system. For this particular system, we consider the downlink shared channel (DL-SCH) that only transmits the physical downlink shared channel (PDSCH). The DL-SCH has many capabilities to increase performance of the system such as multi-antenna systems, beamforming, hybrid automatic repeat request (HARQ), use of demodulation reference signals (DMRS), and use of phase tracking reference signals. 
We consider how a jammer might obtain reward feedback in the form of observed acknowledgment/non-acknowledgement (ACK/NACK) resulting from the victim's attempts to decode the jammed data transmission. Moreover, we consider unreliable observation of this reward information (eg. ACK/NACK) as would be expected with practical detectors.



\begin{figure}[ht]
	\centerline{\includegraphics[width=0.5\columnwidth]{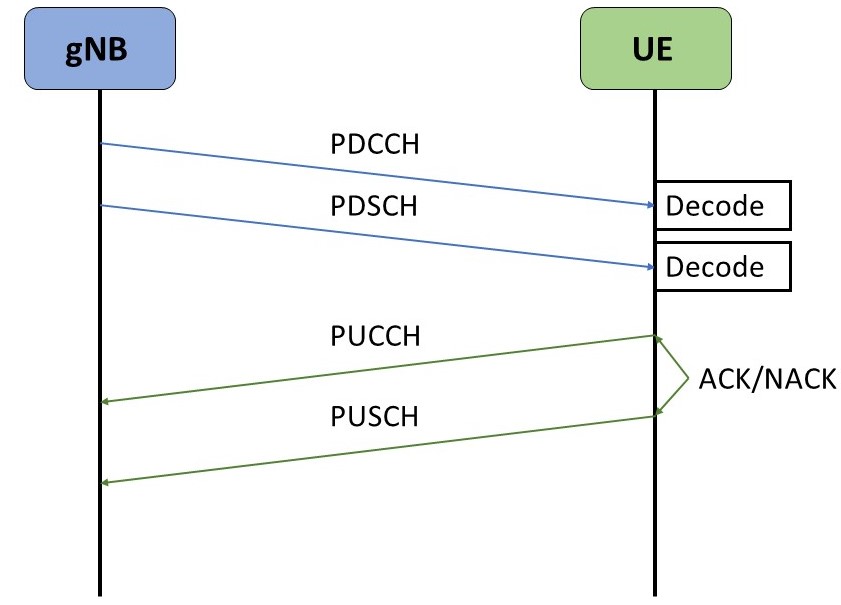}}
	\caption{Overall transmission diagram between a base station (gNB) and a UE.}
	\label{fig:TxDiagram}
\end{figure}

An important question regarding practical implementation of RL for jamming is how the RL agent obtains its reward feedback. In a real life scenario, as shown in Fig. \ref{fig:TxDiagram},
a UE blindly decodes the physical downlink control channel (PDCCH) every millisecond. If decoded, an identifier will point to where the PDSCH is located and also the allocation information for the eventual uplink transmission by the UE. If the PDSCH is successfully decoded, the UE will send an ACK on the uplink transmission either located on the physical uplink control channel (PUCCH) or physcial uplink shared channel (PUSCH). If a block error occurs due to failure to decode the PDSCH, the UE will transmit a NACK. Depending on if HARQ processing is in use, the base station will either retransmit information or not retransmit the information. In the presence of extreme interference or jamming, the UE may not process that the PDCCH was sent. In this case, the UE will not be aware of the need to transmit an ACK/NACK since it did not receive the PDCCH, and thus was not aware of the PDSCH. To recap, a summary of the three scenarios is shown here:
\begin{enumerate}
  \item Miss PDCCH. \(\rightarrow\) No attempt to decode PDSCH \(\rightarrow\) No ACK/NACK transmission. \label{item1}
  \item Detect PDCCH \(\rightarrow\) Detect PDSCH \(\rightarrow\) Transmit ACK. \label{item2}
  \item Detect PDCCH \(\rightarrow\) Fail to decode PDSCH \(\rightarrow\) Transmit NACK. \label{item3}
\end{enumerate}
There are multiple options for the jammer to realistically observe ACK/NACK information. One option is to assume the jammer has knowledge of the victim signal and is able to decode uplink information in order to know whether an ACK or NACK was sent. This adds more complexity to the jammer model to make these observations. An energy detector can also be implemented and used in multiple methods.
In all of the above scenarios, the jammer will receive some partial or noisy observation of the reward. To account for this,
we simply implement the probability of the jammer correctly observing an ACK as \((1-\lambda_{ACK})\) and not observing an ACK as \(\lambda_{ACK}\). This is similarly implemented for NACK observation as well.
The goal is to characterize the impact of the unreliable reward information on the bandit algorithm as a function of the reliability parameter \(\lambda\). 
\section{Learning Problem}
Linear contextual bandits provides a way for the bandit to learn about similar actions from the linear relationships between the distribution of statistics of the action space and the contexts learned from the action space. This is especially important in a continuous action space that is large because it provides a way for the bandit to exploit information between related actions. However, in order to learn features of a particular action, the action still needs to be chosen in order to learn the context vectors of an action set.

We draw from the same linear bandit structure from \cite{Thornton2022,Schutz2024} to conduct this study. The authors from \cite{Thornton2022,Schutz2024} more rigorously define and explain the algorithm used for a full understanding of the learning problem. Here, we summarize key items and review changes made from the learning problems in \cite{Thornton2022,Schutz2024}, such as a change in the action space or assumptions made.

The action space is composed of three components: signaling scheme, probability parameter, \(\rho\), and jamming method. The signaling scheme set is composed AWGN, BPSK, BPSK \(\pi/4\), QPSK, and QPSK \(\pi/4\), where FD jamming is used. The power/jamming probability parameter is \(\rho \in (0, 1]\), and is discretized by M: \(\{1/M,2/M,...,1\}\).
We also consider different jamming methods such as jamming the PDSCH data only, the DMRS only, or randomly within an entire time slot (i.e. both PDSCH data and DMRS).
The instantaneous power provided by the selection of \(\rho\) will be adjusted accordingly depending on the jamming method selected to attain the same average JNR seen by the victim throughout the simulation. This is because when jamming only the PDSCH data or only the DMRS, these only account for a fraction of the subcarrier-symbol block. The power after scaling by \(\rho\) must be adjusted to keep the same average power seen at the victim the same no matter the jamming method selected. 

It is assumed the bandit is able to observe the BLER achieved from the selected strategy from a feedback channel consisting of the victim's ACKs and NACKs. We assume the bandit does not have perfect knowledge of the victim's ACKs and NACKs with there being an associated probability,\((1-\lambda)\), that the bandit correctly observes the ACKs/NACKs within the designated time frame. 
We hope to observe the jammer's performance under this unreliable feedback, as it will effect the cost function and context vector. This will inhibit its ability to learn an effective strategy and its overall learning rate.
The bandit will then update the cost function \(C_t\), which is unknown to the bandit \textit{a priori}. The cost function is expressed as,
\begin{equation}
    C_t = max(BLER_t - BLER_{target},0)/JNR_t
\end{equation}
where \(BLER_t\) is the observed BLER at time step \(t\), \(BLER_{target}\) is the target block error rate, and \(JNR_t\) is the average JNR at time step \(t\). \(JNR_t\) is included in this equation to capture the efficiency of resource usage \cite{Thornton2022}, but for the current case, we use a constant JNR in order to examine the effects of the RL under a fixed average power constraint, as in \cite{Schutz2024}.
 
From the cost function, a context vector is constructed in order to learn about multiple contexts from selecting one action. The context vector is constructed as, 
\begin{align} \label{contextVector}
    \varphi_i(t) &= \notag \\
        &\left[\frac{1}{t}\sum_{l=1}^{t}C_l(a_i),\frac{1}{t}\sum_{l=1}^{t}\mathbbm{1}\{C_l(a_i) > \tau\},\max_{l\leq t}C_l(a_i)\right] 
\end{align}
where \(\mathbbm{1}\) is the indicator function, and \(\tau\) is a threshold selected to indicate whether the victim's communication was disrupted \cite{Thornton2022}. The role of \(\tau\) is to capture the frequency with which an action set produces a non-zero error rate. The other contexts monitor the average cost of the action selected and the largest disruption caused by the jammer. 
From the context vector, the bandit discovers the expected costs from the strategies that are chosen. 
The second feature in the context vector, which tracks the frequency of time steps with non-zero error rates, is particularly important when errors occur infrequently.
For a stochastic linear bandit structure, we use the following relationship, 
\begin{equation}
    C_t(a_i) = \langle\varphi_i,\theta \rangle + \eta_t
\end{equation}
where \(\theta\) is the weighting vector and \(\eta_t\) is a conditionally sub-gaussian random variable conditioned on the jammer's knowledge of the history of costs, actions, and contexts \cite{Thornton2022}. This holds for all \(a \in \mathcal{A}\) and \(t \in \mathbbm{N^+}\) \cite{Thornton2022}. The weighting vector is important because it forms a linear relationship between the actions and contexts gained from the actions chosen. An uninformative prior is constructed using a normal distribution which estimates the weighting vector, \(\theta \sim \mathcal{N}(\hat{\theta},B^{-1})\), which are prior set to uninformative values. This results in finding out information about similar expected costs from similar actions. 
\section{Analysis of Jamming a 5G Signal with Unreliable Feedback} \label{Analysis}
We analyze the addition of linear contextual bandits in a simulation environment.
The victim signal is encoded by a LDPC code with a code rate of 0.54 and modulated with 16QAM. This is the lowest code rate available for 16QAM in the modulation coding scheme (MCS) table. A 15kHz subcarrier spacing is used to transmit the signal. The data of the signal is initially spread over 612 subcarriers in a PDSCH grid. A guard band of length 412 is added to the signal where each side of the data-carrying subcarriers is attached with a guard band length of 206 guard bands. After OFDM-modulation has taken place, cyclic prefixes of either length 80 or 72 are added to each OFDM symbol. The sampling frequency of the signal is 15.36MHz while the bandwidth of the signal is 9.18MHz. The bandit and jammer are assumed non-coherent for these simulations. Each time step, 4 frames that are 10ms long are sent out by the victim. The BLER is averaged for those 4 frames and returned as an observation for the bandit. The value for \(\tau\), the success parameter, was set to 0.5 for all simulations to compare fairly. 
For these simulations, we cumulatively average one simulation between time steps, and then average 20 simulations across time steps.
While averaging the BLER across time steps of multiple simulations provides a better sense of absolute performance of the jammer, we instead observe the average cumulative BLER over multiple simulations because this method allows to better visualize the comparison of different algorithms or scenario parameters such as \(\lambda\). Our interest is in relative performance difference between two results.

We consider an SNR = 24 dB and JNR = 11.2 dB, 7.2 dB, and 6.2 dB. We examine 0.05, 0.1, and 0.15 for values of \(\lambda\). We consider the probability of correctly observing an ACK and the probability of correctly observing a NACK to be symmetrical for this case (eg. \(\lambda = \lambda_{ACK} = \lambda_{NACK}\)). The victim has the ability to use HARQ processing to attempt to increase its throughput and decrease its BLERs.

\begin{figure}[ht]
	\centerline{\includegraphics[width=1\columnwidth]{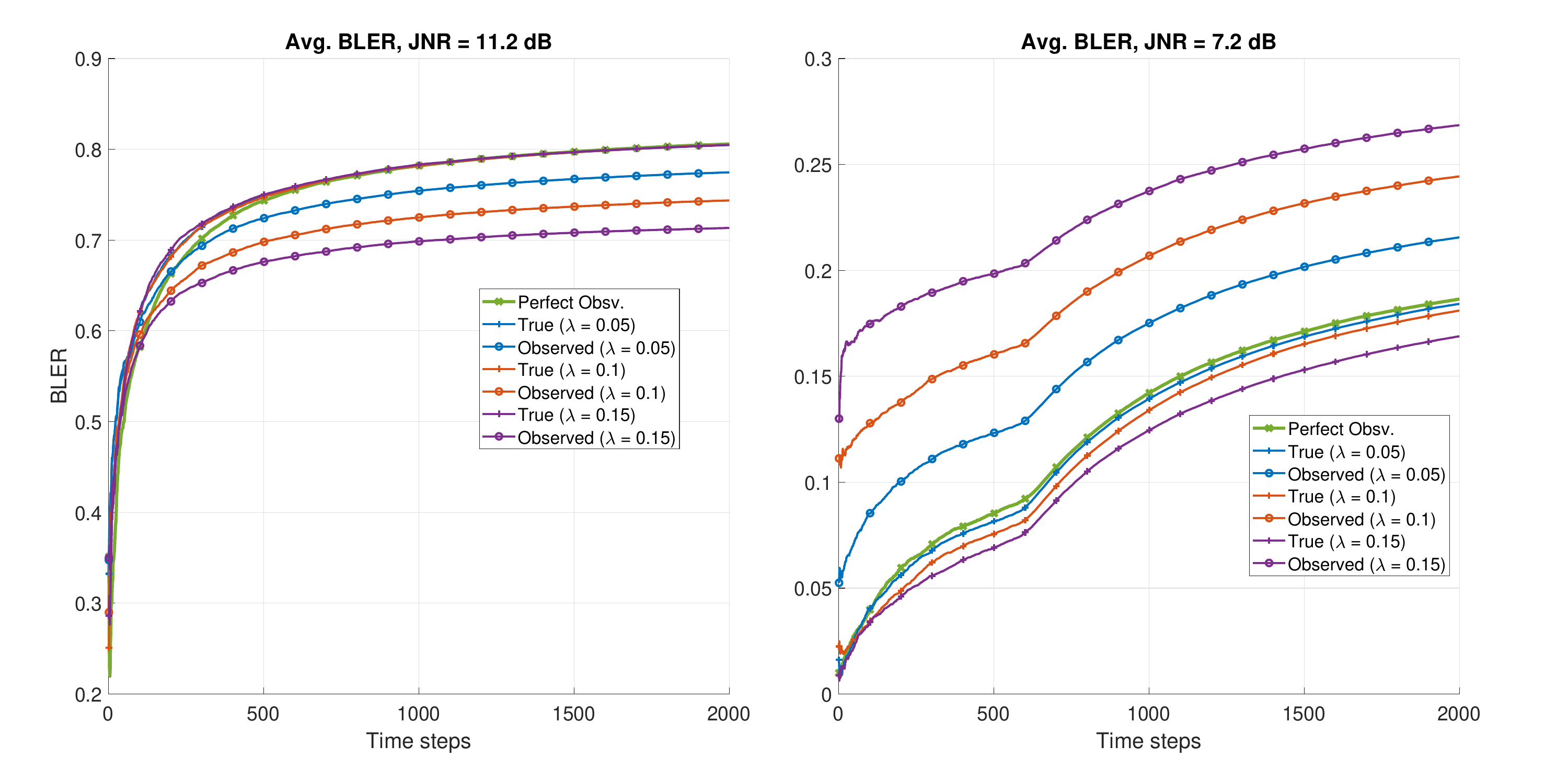}}
	\caption{Comparison of average BLERs between reliability values (\(\lambda\)) at JNRs of 11.2 dB (left) and 7.2 dB (right).}
	\label{fig:BLERcompare}
\end{figure}

Fig. \ref{fig:BLERcompare} shows the BLERs achieved over all the values of \(\lambda\) for the JNRs of 11.2 dB and 7.2 dB. Fig. \ref{fig:BLERcompare} not only shows the BLER the jammer believes it obtained at the different values of \(\lambda\), called the observed BLERs, but also the BLER actually obtained from the actions it took, called the true BLERs, and the BLER obtained by a jammer operating on perfect observations of ACKs/NACKs while non-coherent to the victim signal.
For a JNR of 11.2 dB, the true BLERs achieved from the jammer operating under unreliable information of the ACKs/NACKs is nearly identical to the jammer operating on perfect feedback information for every value of \(\lambda\). They demonstrate short learning periods in order to effectively jam the victim signal. Learning period in this context refers to the actions in which the bandit clearly begins to exploit the victim. The true BLERs also initially perform better than the jammer with perfect observations. This is due to when the jammer achieves a NACK and incorrectly observes that the victim achieved an ACK. This scenario will occur when the SJR is high. Meanwhile, as \(\lambda\) increases, the BLER the jammer believes it achieves decreases. This shows that with high SJR, the performance of the jammer will remain relatively unaffected by the increase in unreliability of the ACK/NACK information, but will believe that its own performance is being degraded. However, at higher values for \(\lambda\), such as \(\lambda = 0.3\), this may change such that the 
BLER caused by the jammer may decrease.

For a JNR of 7.2 dB, the BLER achieved with perfect observations of the ACKs/NACKs are similar to the true BLERs achieved. The true BLERs also decrease as unreliability (\(\lambda\)) increases.
Meanwhile, the observed BLERs by the jammer are much higher than both the true BLERs and the perfect observation BLER. As \(\lambda\) increases, the observed BLERs also increase. This is the result of true NACKs now being infrequent compared to when the SJR is high, so the rewards feedback is now dominated by false NACKs. 
This is opposite of what was observed when the JNR was 11.2 dB.
The learning rates have also increased significantly compared to when the JNR was 11.2 dB.

\begin{figure}[ht]
	\centerline{\includegraphics[width=1.0\columnwidth]{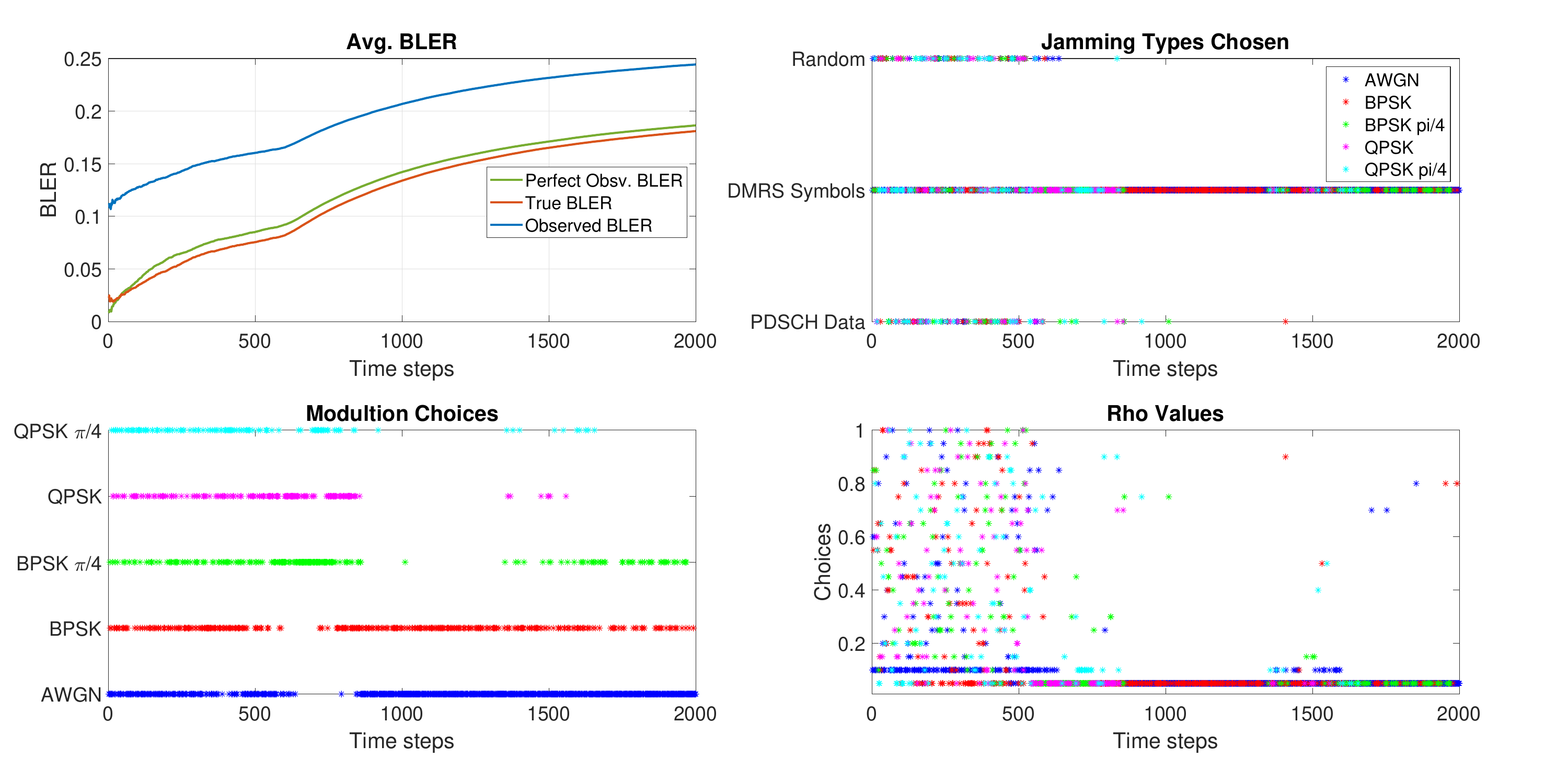}}
	\caption{Collective results under HARQ processing and unreliable feedback (\(\lambda = 0.1\)) of BLER, types of jamming methods used, jamming modulation schemes used, and \(\rho\) values chosen at JNR = 7.2 dB and SNR = 24 dB.}
	\label{fig:detailedJNR}
\end{figure}

Fig. \ref{fig:detailedJNR} shows the BLER achieved and choices the jammer made when the JNR \( = 7.2\) dB and \(\lambda = 0.1\). 
The jammer clearly found that jamming the DMRS led to higher BLERs because it gave the victim receiver incorrect channel and noise estimation factors severely inhibiting equalization of the incoming signal. The BLER examination is important for this section because it shows the convergence rate and how high it converges to a particular BLER. However, the strategies the jammer chooses is also of interest because it shows how the jammer learned to reach that point. 
There is a margin of error between the true BLER and observed BLER that the jammer believes it caused. 
As previously stated this is because true NACKs are now infrequent compared to when the SJR is high, so the rewards feedback is dominated by false NACKs. This leads to the observed BLER being larger than the true BLER.  The main modulation choices used are BPSK and AWGN with partial usage of BPSK \(\pi/4\). The transition period between learning period and exploiting period is clear in these plots where there is an obvious transition at ~700 time steps. This means the jammer is still able to learn, but is more unsure of its choices because of the unreliability of the reward function. It will eventually converge to a BLER that will heavily impact the victim receiver (10\% BLER in about 28 seconds). The main aspect is that it can still differentiate between jamming methods to effectively jam the victim by choosing DMRS jamming.

For a JNR of 6.2 dB, the jamming power was too low to increase the BLERs to at least 2\% block errors. Either the power will need to be increased if there are no power constraints on the jammer, or some properties of the bandit may need to be modified, such as the context vector.
\section{Conclusion and Future Work}
Jamming a coded system introduces new complexities in choosing an effective jamming scheme.
The jammer must try to distribute the LLRs in a way that is irrecoverable for the error correction code. From simulations, we have demonstrated that targeting whole codewords is the best way to do this. Depending on the SJR however, how the power is distributed in the codeword is important. 

When unreliable reward feedback is introduced to the bandit, the jammer's performance and ability to learn degrades significantly depending on the JNR. With high reliability (i.e. 95\% correct ACK/NACK observation), the jammer is still able to learn to significantly effect the victim. When reliability is reduced, the jammer's learning capabilities are reduced, but if the SJR is small, it can overcome the unreliability. If the SJR is large, the jammer may need to increase power to effectively learn to jam the victim. An alternative if there are power constraints would be to manually modify the bandit, such as the context vector, to improve performance.


Modern communications systems will also lower MCS if degradation in the system is observed. From our results, once a BLER of 10\% is noticed by the victim signal the victim would have switched to a more robust modulation scheme and coding rate, such as QPSK with LDPC code rate of 0.54. A victim lowering its MCS could be considered a success from the jammer's perspective. However, the jammer will likely see a decrease in BLER as a result which will impact the ability to learn successful jamming strategies. Non-stationary tracking of the problem is important to discover performance of the jammer employing a RL algorithm for implementation in real-life scenarios.

\bibliographystyle{IEEEtran}
\bibliography{IEEEabrv,LinearJammingBandits}
\nocite{*}
\end{document}